\documentstyle[11pt,twoside,newpasp,psfig,epsf]{article}

\newcommand{\gtap}{\mathrel{\hbox{\rlap{\lower.55ex \hbox {$\sim$}}
                   \kern-.3em \raise.4ex \hbox{$>$}}}}
\newcommand{\ltap}{\mathrel{\hbox{\rlap{\lower.55ex \hbox {$\sim$}}
                   \kern-.3em \raise.4ex \hbox{$<$}}}}

\begin{document}

\title{Binary evolution and neutron stars in globular clusters}

\author{Frank Verbunt}
\affil{Astronomical Institute, Utrecht University, Postbox 80.000,
   3508 TA Utrecht, The Netherlands; email verbunt@phys.uu.nl}

\begin{abstract}
Improved observations of globular clusters are uncovering a large number
of radio pulsars and of X-ray sources. The latter  
include binaries in which a neutron star or a white dwarf accretes matter
from a companion, recycled pulsars, and magnetically active binaries.
Most of these sources originate from close encounters between stars in
the cluster core; magnetically active binaries and some cataclysmic
variables have evolved from primordial binaries. The formation rate
through close encounters scales differently with the central density and core
radius of the cluster than the probability for a single binary to
be perturbed by an encounter. This is exploited in some preliminary
observational tests of the close encounter hypothesis. \\
Accreting black holes have been found in globular clusters with other
galaxies; the absence of such holes in the Milky Way
clusters is compatible with the small number expected.
\end{abstract}


\section{Introduction}

\begin{figure}[]
\centerline{
\parbox[b]{8.0cm}{\psfig{figure=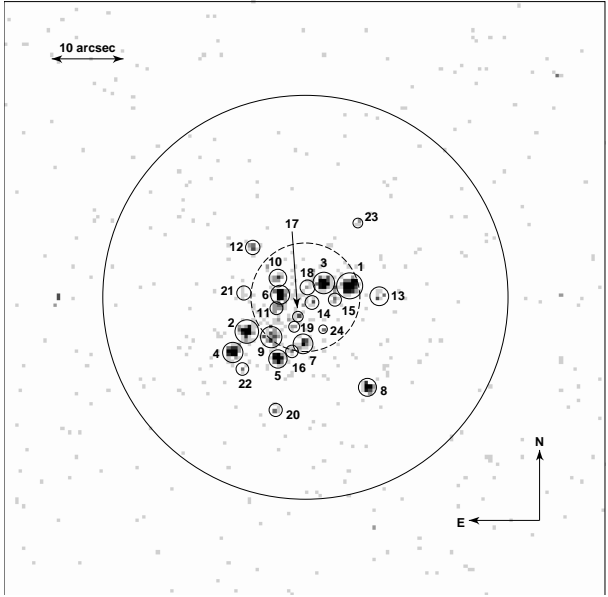,width=8.cm,clip=t}}
\parbox[b]{6.0cm}{\caption{
X-ray sources in NGC6440. Sources 1 certainly and 2, 3 and 5 probably are 
neutron stars accreting from a companion star at a low rate; most of the
other sources are probably white dwarfs accreting from a companion,
i.e.\ cataclysmic variables. Note that the source distribution extends
beyond the core radius (indicated with a dashed circle), but remains well 
within the half-mass radius (solid-line circle). From Pooley et al.\ (2002b).
\label{verbuntngc}}}}
\end{figure}

Interest in binaries in globular clusters revived when the first X-ray
maps of the whole sky showed an overabundance in globular clusters
of bright ($L_x\gtap10^{36}$ erg/s) sources: whereas globular clusters
contain $\sim$0.1\%\ of the stars of our galaxy, they contain about
10\%\ of the bright X-ray sources. 
It was soon suggested that close stellar encounters in globular cluster
cores were responsible (Clark 1975).
With more sensitive instruments faint  ($L_x\ltap10^{35}$ erg/s) sources
were detected in globular clusters: 8 in 8 clusters with the Einstein
satellite in the 1980s (Hertz \&\ Grindlay 1983), and 57 in 23 clusters 
with ROSAT in the 1990s (Verbunt 2001).
Of the latter 57, 17 are more than two core radii from
the cluster core. 
In the new millenium, Chandra detected about 100 faint X-ray sources in 47 Tuc
alone, and 25-40 sources in each of NGC6752, NGC6397, and NGC6440 
(Grindlay et al.\ 2001a,b, Pooley et al.\ 2002a,b;
Fig.\,\ref{verbuntngc}).
First XMM results are now being published (Webb et al.\ 2002; Webb, Gendre in
these proceedings).

Radio pulsars have also been detected in large numbers: 20 in 47 Tuc alone,
8 in M15, 5 in NGC6752, and some 20 in a dozen other clusters (Freire et al.\
2001, D'Amico et al.\ 2002, and lists in e.g.\ Phinney 1992, Lyne 1994, and
on the Web: Freire 2002).
Almost all of these are recycled radio pulsars, i.e.\ have obtained their
rapid rotation (and probably also their low magnetic field) through
accretion of matter and angular momentum from a companion star.

At the moment 13 bright X-ray sources are known in a total of 12 globular
clusters. The only binaries producing such luminosities are low-mass
X-ray binaries, in which a neutron star or black hole accretes mass from
a companion. Because of the occurrence of X-ray bursts, 12 of the 13
bright sources in clusters are known to be neutron stars; the 13th one
is -- on the basis of its X-ray spectrum -- probably a neutron star as
well (in 't Zand et al.\ 1999, White \&\ Angelini 2001). 
Thus there is no evidence for accreting black holes in any globular
cluster with our galaxy.
The dim sources with $10^{32.5}\ltap L_{x}{\rm (erg/s)}\ltap10^{35}$
are most likely neutron stars accreting at a low rate; most of the
dimmer sources ($\ltap10^{32.5}$) are probably cataclysmic variables,
but some of them are recycled radio pulsars, or magnetically active
close binaries (`RS CVns') -- see e.g.\ Fig.8 in Verbunt et al.\ (1997);
and the articles by Cool and Edmonds in these proceedings.

In this review I discuss the formation and evolution of these X-ray sources
and recycled radio pulsars. Formation through evolution from a primordial
binary is compared with formation via close stellar
encounters in Section\,2. The evolution of binaries with a compact star
(neutron star or white dwarf) is
described in Section\,3. Various tests of our theoretical ideas
are made by comparison with observation in Section\,4. A brief discussion
in Section\,5 of X-ray sources in globular clusters of other galaxies,
precedes the Summary.

\section{Formation: evolution versus encounters}

All binaries that we observe in the disk of the galaxy have evolved from
primordial binaries. 
Binaries which are very common in the disk, apparently are
commonly formed via ordinary binary evolution. If we find such binaries
in a globular cluster, it is likely that they have evolved
from primordial binaries as well.
Examples of such binaries are RS CVn's and contact binaries.
On the other hand, binaries with neutron stars (or black holes)
are extremely rare in the disk; such a binary in a globular cluster
is formed almost certainly from a close encounter of a neutron star
with a single star or with a binary.
Cataclysmic variables are in between these extremes: in globular clusters
with dense cores, such as 47 Tuc, most may have formed via stellar encounters;
but in more open clusters, as $\omega$ Cen, they may have evolved from
primordial binaries (Verbunt \&\ Meylan 1988).

We consider tidal capture first. The encounter rate is proportional
to the encounter cross section $A$, to the relative velocity $v$,
and to the numbers $n_c$ and $n$ per unit volume, of compact stars and 
ordinary stars respectively. 
To obtain the encounter rate $\Gamma$ for the cluster
as a whole, one integrates over the cluster volume; because of
the high densities in the core this roughly corresponds to multiplying
the central rate with the core volume.
Due to gravitational focussing, the cross section $A$ of close encounters
is proportional to the radius of the star $R$
and inversely proportional to the square of the velocities $v$
(see also Davies, these proceedings). Thus (Hut \&\ Verbunt 1983):
\begin{equation}
\Gamma \propto \int n_cnAv dV \propto \int {n_cnR\over v} dV
 \propto {\rho_o^2r_c^3\over v}\,R
\label{gamma}
\end{equation}
Where $\rho_o$ is the central mass density and $r_c$ the core radius.
An analogous reasoning gives the exchange encounter rate
\begin{equation}
\Gamma_e \propto \int n_cn_bA_bv dV \propto \int {n_cn_ba\over v} dV
 \propto {\rho_o^2r_c^3\over v}\,a
\label{gamme}
\end{equation}
where $n_b$ is the number of binaries per unit volume, and $a$
the semi-major axis of the binary.
The ratio of tidal capture to exchange encounters is roughly
\begin{equation}
{\Gamma\over\Gamma_e} \sim {R\over a} {n\over n_b}
\end{equation}

The susceptibility of large binaries to close encounters with other
cluster stars implies that many such binaries are dissolved by passing
stars. As a result, the formation of cataclysmic variables from evolution
of primordial binaries, which passes through a stage in which the binary
is very wide, is suppressed in globular clusters (Davies 1997).

The importance of tidal capture is under debate, because it leads to
an initially highly eccentric orbit: the circularization of this orbit
is accompanied by dissipation of an amount of energy $\Delta E$
which is comparable to the binding energy $E_b$ of the ordinary star:
\begin{equation}
{\Delta E\over E_b} \simeq
\left({-GMm\over 2a_c}\right)/\left({3GM^2\over 5R}\right) \simeq
{5\over 6}{m\over M}{R\over a_c}
\end{equation}
where $M$ and $R$ are the mass and radius of the tidally distorted star,
$m$ the mass of the other star and $a_c$ the radius of the circularized orbit.
If the energy is dissipated more rapidly than it can be radiated
or convected away, the ordinary star is destroyed, and no binary remains
(Ray et al.\ 1987).

Tidal capture, if successful, tends to lead to orbits with $a\simeq 3R$, 
i.e.\ to orbits with short periods, less than 1 day, say.
In contrast, exchange encounters favour wide orbits, and in addition
will cause a recoil velocity of the binary which may take it out of
the cluster core -- perhaps even out of the cluster!
Tidal capture can also occur during an encounter of a single star with
a binary, when the three-body intraction brings two stars close
to one another. 

A much debated point of uncertainty is the number of neutron
stars that remain in a cluster after their formation: young pulsars
obtain a kick velocity at birth, as witnessed by the observed velocities
of radio pulsars. It has been argued that these velocities
are higher than the escape velocities of globular clusters, implying
that no neutron stars would be retained. In my opinion,
the velocities of pulsars have been rather overestimated, due
to underestimates of the errors in proper motions and distances
(see Hartman 1997). A recent list of 16 accurate velocities obtained with
VLBI (Brisken et al. 2002) indicates that as much as a third of neutron 
stars is born with velocities less than $\sim 50$\,km/s.
Thus as many as 30\%\ of the neutron stars born in globular
clusters are retained.
Loss of neutron stars from globular clusters is counteracted by
mass segregation, which concentrates the (relatively heavy) neutron stars
in the cores, where the encounter rates are highest (Verbunt \&\ Meylan
1988).

The presence of pulsars in globular clusters is interesting, as they
constrain the amount of non-luminous matter (to which they themselves
do not contribute significantly, as the mass in white dwarfs is always 
much higher). A pulsar which has a radial velocity $v_r$ with respect
to us, will be observed at a period shifted by $\Delta P=(v_r/c)P$.
If the pulsar is accelerated, a period derivative $\Delta\dot P = (a_r/c)P$
is measured on top of the intrinsic period derivative. Since the
latter is always positive, measurement of a negative period derivative
indicates that acceleration dominates, and thus provides an estimate
of the mass density. Phinney (1992) has shown that the pulsars in M15
put a tight constraint on the mass of any black hole in the cluster
core -- the most likely mass of the cluster core can in fact be explained
by the sum of visible stars and white dwarfs, if the mass distribution
initially followed the Salpeter function, i.e.\ no black hole is
required.

\section{Binary evolution}

The evolution of a binary with a neutron star (or black hole, or
white dwarf) is a complex topic, which is described only briefly
here. For more detail, see e.g.\ Verbunt (1993).
The mass transfer in a close binary in which a companion to a neutron
star fills its Roche lobe, can be driven by loss of angular momentum
from the orbit. If a compact star of mass $m$ receives matter from
a Roche-filling star with mass $M$, the mass
transfer rate is roughly given by $-\dot M/M\simeq \dot m/m\sim \dot J/J_b$,
where $J_b$ is the binary angular momentum.
For a main sequence star donor, the orbital period $P_b$ is roughly given
by $P_b{\rm (hr)} \simeq 8 M(M_{\odot})$, and gravitational radiation
leads to a mass transfer rate of order $10^{-10}M_{\odot}\,{\rm yr}^{-1}$.
When accreted onto a neutron star this leads to an X-ray luminosity
$\sim 10^{36}$ erg\,s$^{-1}$.
The observation of higher X-ray luminosities in such binaries have lead
to the suggestion of additional mechanisms of loss of angular momentum,
such as 'magnetic braking', in which a stellar wind carries away
angular momentum efficiently because it is tied to the stellar
surface out to large radii by  magnetic field lines. The efficiency
of magnetic braking is unknown; recent studies of stellar rotation
in young star clusters suggests that its importance may have been
over-estimated (e.g.\ Tinker et al.\ 2002).
The mass transfer rate may also be affected if the donor is irradiated
by X-rays from the accreting star. If this irradiation causes the star
to expand, it may enhance the mass transfer rate; if it causes an
enhanced stellar wind and mass loss from the binary, it may
reduce the mass transfer rate. Whether irradition is important
is not known; the absence of strong heating effects in most 
well-observed low-mass X-ray binaries suggests that the accretion
disk shield the secondary from X-rays emitted near the accreting star.

If the mass transfer stops, the neutron star may spin rapidly enough
to switch on as a radio pulsar. The standard description of evolution
through loss of angular momentum does not foresee such an end however,
but rather predicts an everlasting, albeit ever lower, mass transfer.
The observation of close binaries with a pulsar shows that the mass
transfer does, in fact, stop. Perhaps if the mass transfer is variable,
it may reach a state sufficiently low that the pulsar switches on, after
which the pulsar wind itself can prevent further mass transfer.
Once more, the details of such a process are not understood.

The star that donates matter to a neutron star can also be a white dwarf;
in that case the orbital period is roughly 
$P_b{\rm (s)}\simeq 50/M(M_\odot)$. In such a binary, stable mass
tranfer can only occur if the white dwarf has a mass less than 
$\sim0.66M_{\odot}$; a more massive white dwarf will expand so rapidly 
with mass loss that it is disrupted within one binary revolution 
(Verbunt \&\ Rappaport 1988).
 
If the initial binary is too wide, loss of angular momentum is not important,
and mass transfer will only start when the donor expands 
as it ascends the (sub)giant branch.
In that case $-\dot M/M\simeq \dot m/m\sim\dot R/R$, where $R$ is
the stellar radius. According to stellar evolution, $\dot R$ increases
with $R$, and thus the mass transfer rate is expected to be higher
in wide binaries.
Conservation of angular momentum widens the orbit as mass is transferred
from the (sub)giant to the neutron star; typically, the orbital period
has increased by a factor $\sim7$ by the time that the whole envelope
has been transferred. From that point on, the binary will consist of
a neutron star and a white dwarf.
The mass transfer puts an early end to the growth of the giant core,
and the emerging white dwarf is undermassive; typically 
$M_{wd}\sim0.25M_{\odot}$. Tidal interaction during the mass transfer
leads to an orbit which is almost but not quite circular.
The observed white dwarf masses and the relation between orbital period
and (small) eccentricity provide strong support for this evolution
scenario for recycled pulsars with white dwarf companions (Phinney 1992).

It is observed that rather more millisecond pulsars are found in globular
clusters than can be formed from the currently observed X-ray binaries.
A solution could be that most recycled pulsars were spun up in binaries
with donors of intermediate mass, 1-3 $M_{\odot}$ (Davies \&\ Hansen 1998).

\section{Formation through close encounters: testing the theory}

\begin{figure}[]
\centerline{\psfig{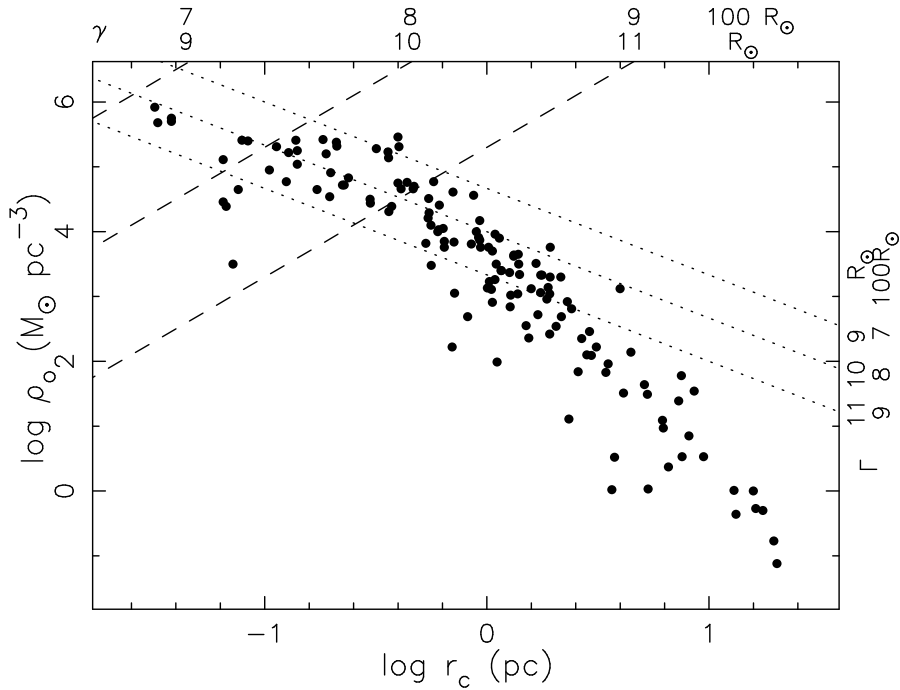}}
\caption{Central density as a function of core radius
for globular clusters in our Galaxy. Data from Harris (1996,
revision of June 22 1999). Dotted lines
indicate loci of constant formation rate $\Gamma$ in the cluster, according
to Eq.\,\ref{gammc}; approximate numerical factors are indicated on the
right for $R=R_{\odot}$ (applicable to tidal capture) and for 
$a=100 R_{\odot}$, $n_b\simeq n$
(exchange encounters) where $x$ indicates one encounter per $10^x$ yr.
Dashed lines indicate lines of constant encounter rate $\gamma$
for {\em one single} system, according to Eq.\,\ref{gammd}; approximate 
numerical factors are indicated on top.
\label{verbuntrhc}}
\end{figure}

With a growing number of binaries with neutron stars  observed in globular
clusters, we can start comparing their observed properties with the
theory of formation. 
To do so, we refer to Eqs.\,\ref{gamma} and \ref{gamme}, and note
that the virial theorem relates the velocity dispersion to the
central density and core radius of the cluster:
\begin{equation}
v\propto \sqrt{\rho_o}\,r_c
\label{vel}\end{equation}
Therefore
\begin{equation}
\Gamma \propto {\rho_o}^{1.5}{r_c}^2R \qquad {\rm and} \qquad
\Gamma_e \propto {\rho_o}^{1.5}{r_c}^2a
\label{gammc}\end{equation}
In Figure\,\ref{verbuntrhc} we plot lines of constant $\Gamma$ in a graph
showing the central density as function of core radius for the globular
clusters of our Milky Way.
The numbers indicated use the King model for the numerical factor
in Eq.\,\ref{vel}.
The formation rate of binaries with neutron stars can be high 
because the central density is high, or because the cluster has a 
large core.

The situation is different for the probability that a binary, once
formed, will undergo a subsequent encounter which may change or destroy it.
The rate at which a binary undergoes close encounters is given by
\begin{equation}
\gamma = nAv \propto {\rho_o\over v}\,a \propto {{\rho_o}^{0.5}\over r_c}\,a
\label{gammd}\end{equation}
Lines of constant $\gamma$ are also shown in Figure\,\ref{verbuntrhc}.
We see that in the clusters with the highest central density,
both close binaries (formed from tidal capture) and wider binaries
(formed from an exchange encounter) are affected by subsequent encounters.

\begin{figure}[]
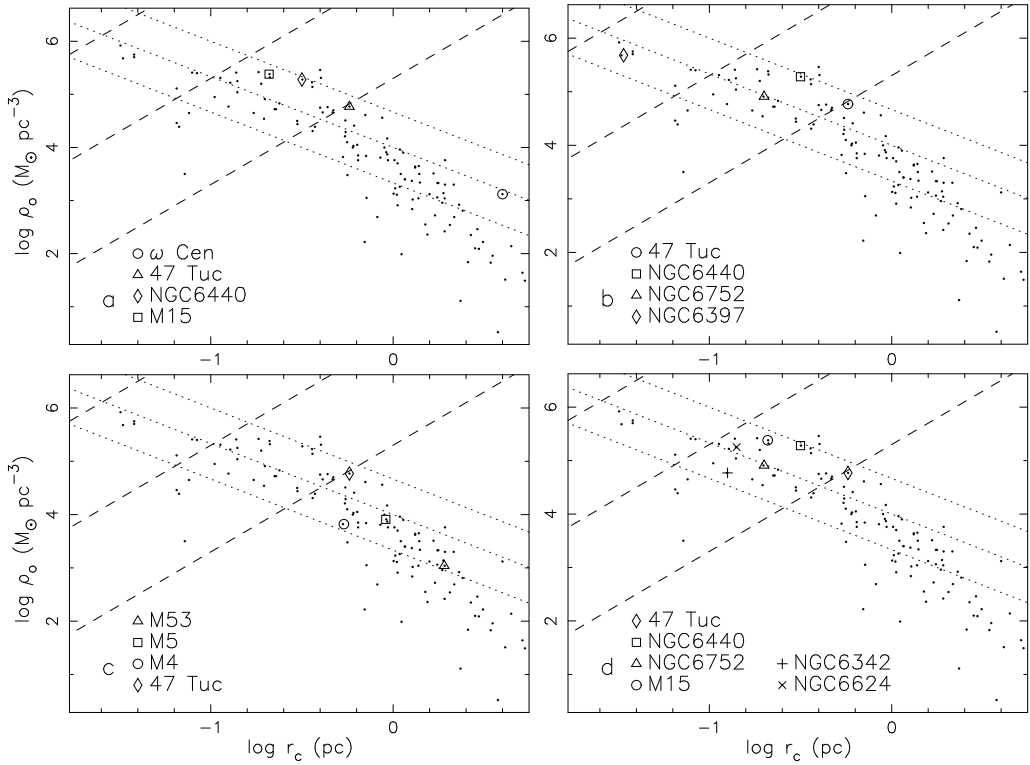

\centerline{
\parbox[b]{7.0cm}{\psfig{figure=verbuntf3a.ps,width=7.cm,clip=t}}
\parbox[b]{6.5cm}{\psfig{figure=verbuntf3b.ps,width=6.5cm,clip=t}}
}
\centerline{
\parbox[b]{7.0cm}{\psfig{figure=verbuntf3c.ps,width=7.cm,clip=t}}
\parbox[b]{6.5cm}{\psfig{figure=verbuntf3d.ps,width=6.5cm,clip=t}}
}
\caption{Various -- preliminary -- tests on neutron stars in binaries
in globular clusters.  a) Number of binaries with neutron stars, b) 
Slope of the luminosity function of the X-ray sources,
c) Orbital periods of pulsar binaries,
d) Pulsar pulse periods. For details see text.
\label{verbuntf3}}
\end{figure}

The exact encounter rates in a cluster depend on the
mass function in the core, hence on the mass segregation,
on the fraction of stars in binaries, and on the period
distribution of the binaries. For encounters involving
neutron stars, the rates depend on the retention fraction of neutron stars.
Also, it is not clear how one should treat a collapsed cluster.
All these factors are different in different clusters.
In the absence of detailed information on most clusters,
we can only perform preliminary tests, in which we ignore
these factors.

One such test is that the probability that a cluster 
contains a bright X-ray source scales with $\Gamma$, i.e.\ with
the collision number $\Sigma\equiv\rho_c^2r_c^3/v$. The bright
X-ray sources pass this test (Verbunt \&\ Hut 1987). 
Johnston et al.\ (1992) claim to find that
the probability that a cluster contains a recycled pulsar
is less dependent on $\rho_c$, viz. $\propto\rho_c^{1.5}$; however,
this lower dependence may be an artefact of their assumption that
$v$ is the same in all clusters (compare eqs.\ref{gamma} and \ref{gammc}).

Some further tests are illustrated in Figure\,\ref{verbuntf3}.
Numbers used in these tests for the X-ray sources / radio pulsars 
were taken from: $\omega$\,Cen Rutledge et al.\ 2002 / --;
47\,Tuc Grindlay et al.\ 2001a / Freire et al.\ 2001; NGC6440 Pooley et 
al.\ 2002b / Lyne et al.\ 1996; M15 White \&\ Angelini 2001 / Phinney 1992;
NGC6752 Pooley et al.\ 2001a / D'Amico et al. 2002; NGC6397 Grindlay et 
al.\ 2002b / --. Numbers for other pulsars from: M53 Kulkarni et al.\ 1991; 
M5 Anderson et al.\ 1997; M4 Thorsett et al.\ 1999; NGC6342 Lyne et al.\ 1993;
NGC6624 Biggs et al.\ 1994. 

{\bf a}) $\omega$\,Cen contains one neutron star X-ray binary
and no (known) pulsar, whereas 47\,Tuc, NGC6440 and M\,15
contain respectively 2, 4 and 2 neutron star X-ray binaries
and 20, 1 and 8 (known) pulsars. $\omega$\,Cen indeed has a lower
value of $\Gamma$ than the other three clusters.

{\bf b}) The slope $p$ of the X-ray luminosity function
$dN(L_{\rm x})\propto {L_{\rm x}}^{-p}d\ln L_{\rm x}$ (including 
neutron star binaries, cataclysmic variables, pulsars, and
magnetically active binaries) is steep in 47\,Tuc ($p=0.8$),
intermediate in NGC6752 and NGC6440 ($p=0.5$), and shallow in
NGC6397 ($p=0.3$). This may be related to the high value of $\gamma$ in
NGC6397, which prevents a binary from evolving without being
interfered with; thus ordinary binaries in this cluster
have largely been destroyed (Pooley et al.\ 2002b).

{\bf c}) The clusters M4 and M53 contain pulsars in binaries with
long orbital periods of 191 and 256 days; these clusters indeed have
a low value of $\gamma$, i.e.\ wide binaries are not affected by
encounters. 47\,Tuc, with a higher value for $\gamma$, does
not contain pulsars in binaries with periods longer than 2.3\,d.
(The binary in M4 in fact has a companion third star at about
6000\,$R_{\odot}$; the outer binary has only an expected life time
of $10^8$\,yr.)

{\bf d}) It has been an outstanding puzzle that all pulsars in 47\,Tuc
and NGC6752 have very short pulse periods ($<7.6$\,ms
and $<9.0$\,ms), whereas those in M15 have
pulse periods ranging up to 111\,ms. Long periods also are found
in NGC6342, NGC6440 and NGC6624 (1.0\,s, 289\,ms and 379\,ms).
There is no obvious correlation of the pulse period range with 
$\Gamma$ or $\gamma$.

The encounter hypothesis for the formation 
of binaries with a neutron star survives the preliminary tests.

\section{Globular clusters in other galaxies}

With the Einstein and Rosat satellites, X-ray sources in globular clusters
of M31 have been discovered. On the whole it appears that the properties
of these sources are not significantly different from those in our
galaxy (Supper et al.\ 1997). 
Thanks to Chandra it has become possible to observe X-ray sources
in other galaxies, with fairly accurate positions. As a result, it
has been found that globular clusters in other galaxies contain X-ray
sources with luminosities above $10^{39}$\,erg/s (Angelini et al.\ 2001,
Sarazin et al.\ 2001, Kundu et al.\ 2002)
At such luminosities, the X-ray sources are probably black holes
accreting from a binary companion; this is confirmed by the soft
X-ray spectrum when it can be measured.
Apparently the clusters in those galaxies {\it do} contain black holes
in binaries. 
This is not necessarily a significant difference
with the clusters in our galaxy: if 20 \%\ of the bright X-ray
sources in globular clusters would contain black holes, the expectation value
for such binaries in our globular cluster system would be 2 or 3. For this
expectation value the probability of finding zero is appreciable.

A first contribution by XMM is the detection of a faint X-ray source
in Mayall II, the largest cluster of M31 (Verbunt, Meylan \&\ Mendez,
in preparation).

\section{Summary}

The overabundance in globular clusters of binaries with neutron stars is
due to formation of such binaries in close encounters. 
When the neutron star accretes matter from its companion, it is
an X-ray source. Chandra observations show that there are 5 to 10 times
as many neutron stars accreting at low rates than at high rates.
Whether this is related to their formation mechanism is not clear,
since a similar overabundance is found in the galactic disk where
binaries with neutron stars evolve from primordial binaries
(Cornelisse et al.\ 2002).
When the neutron star stops accreting, it can become a radio pulsar.
Most cataclysmic variables in globular clusters are also formed in close
encounters.

Tidal capture leads to a short orbital period, and has no associated
recoil velocity; exchange encounters favour wide binaries and do
involve recoil.
Most neutron star binaries are found in or close to the cores of globular
clusters. Most binaries with a pulsar have relatively short orbital
periods, and mass functions that indicate a mass of $\simeq0.25M_{\odot}$
for their white dwarf companion. In my view, this indicates that 
these binaries were formed via tidal capture. 
Some binaries have rather longer orbital periods, and some are far
from the cluster core. These binaries were probably formed via exchange
encounters.
From the X-ray luminosity function it appears that magnetically active
binaries -- that make up the low--luminosity end of the distribution --
are destroyed in dense clusters.

Observations of globular clusters in other galaxies show that black holes
are present in them. It would be interesting to investigate whether
any of the low--luminosity X-ray binaries in the globular clusters of
our galaxy contains a black hole.
The small number of X-ray sources in clusters of our Galaxy hampers
comparison of their properties with those in globular clusters systems
of other galaxies.

\end{document}